\documentclass[pre,twocolumn,showpacs,amsmath,amssymb]{revtex4}
\usepackage[dvips]{graphicx}
\usepackage{epsfig}
\input epsf
\usepackage{hyperref}
\usepackage{breakurl}
\usepackage{color}

\begin{document}

\title{Direct experimental observation of binary agglomerates in complex plasmas}

\author{M. Chaudhuri$^*\footnotetext{* corresponding author: chaudhuri@mpe.mpg.de}$, V. Nosenko, C. Knapek, U. Konopka, A. V. Ivlev, H. M. Thomas and G. E. Morfill}

\affiliation{Max-Planck-Institut f\"ur extraterrestrische Physik, 85741 Garching, Germany}

\begin{abstract}

A defocusing imaging technique has been used as a diagnostic to identify binary agglomerates (dimers) in complex plasmas. Quasi-two-dimensional plasma crystal consisting of monodisperse spheres and binary agglomerates has been created where the agglomerated particles levitate just below the spherical particles without forming vertical pairs. Unlike spherical particles, the defocused images of binary agglomerates show distinct, stationary/periodically rotating interference fringe patterns. The results can be of fundamental importance for future experiments on complex plasmas.

\end{abstract}


\maketitle

In recent times with the advancement of technology, it is possible to explore the highly ordered crystalline structures in different domains of nature, ranging from nanotechnology (graphene~\cite{Novoselov_science}, electrons on a liquid helium surface~\cite{Grimes_PhysRevLett.42.795}, etc.), biophysical systems (protein crystal~\cite{Kendrew_science}), photonic crystals~\cite{Yablonovitch_PhysRevLett.63.1950} to soft matter systems (colloids~\cite{Murray_PhysRevB.42.688}, granular material~\cite{Reis_PhysRevLett.96.258001}, etc.). The discovery of plasma crystal and liquids has brought the strongly coupled complex plasmas in the existing hierarchy of the soft matter systems~\cite{Thomas_PRL,Chu_PRL,Thomas_Nature,Morfill_RMP,Chaudhuri_SM}. Complex plasma is a mixture of electrons, ions, highly charged micro-particles and neutral gas.
Due to higher spatial ($\sim \mu$m) and temporal ($\sim$ Hz) scale lengths of individual particle dynamics in the dilute background plasma, complex plasma is used as a model system in soft matter 
to explore  fundamental physics beyond the limits of hydrodynamics (continuous media) and study various generic processes occuring in solids and liquids, in regimes ranging from the onset of cooperative phenomena to large strongly coupled systems at the most fundamental kinetic level~\cite{Fortov2005PR,Morfill_RMP,Shukla_RevModPhys.81.25}.
In this respect, the two dimensional plasma crystal has been used extensively to investigate various generic processes associated with strong coupling phenomena~\cite{Quinn_PhysRevE.64.051404,Nosenko_PhysRevLett.103.015001,Nunomura_PhysRevLett.95.025003,Nosenko_PhysRevLett.93.155004,Nosenko_PhysRevLett.99.025002,Knapek_PhysRevLett.98.015004}. 
Traditionally, the two-dimensional plasma crystal is formed with monodisperse spherical particles after a ``purification process'' through which all the bigger/asymmetric particles are removed. 
Then the conventional video microscopy technique is used to investigate all the static and dynamic properties of plasma crystal and liquid. All the phenomena investigated so far have been analyzed by taking focused images of dust particles and then identifying their positions and tracking their motion accurately. 

\begin{figure}
\includegraphics[width=0.85\linewidth]{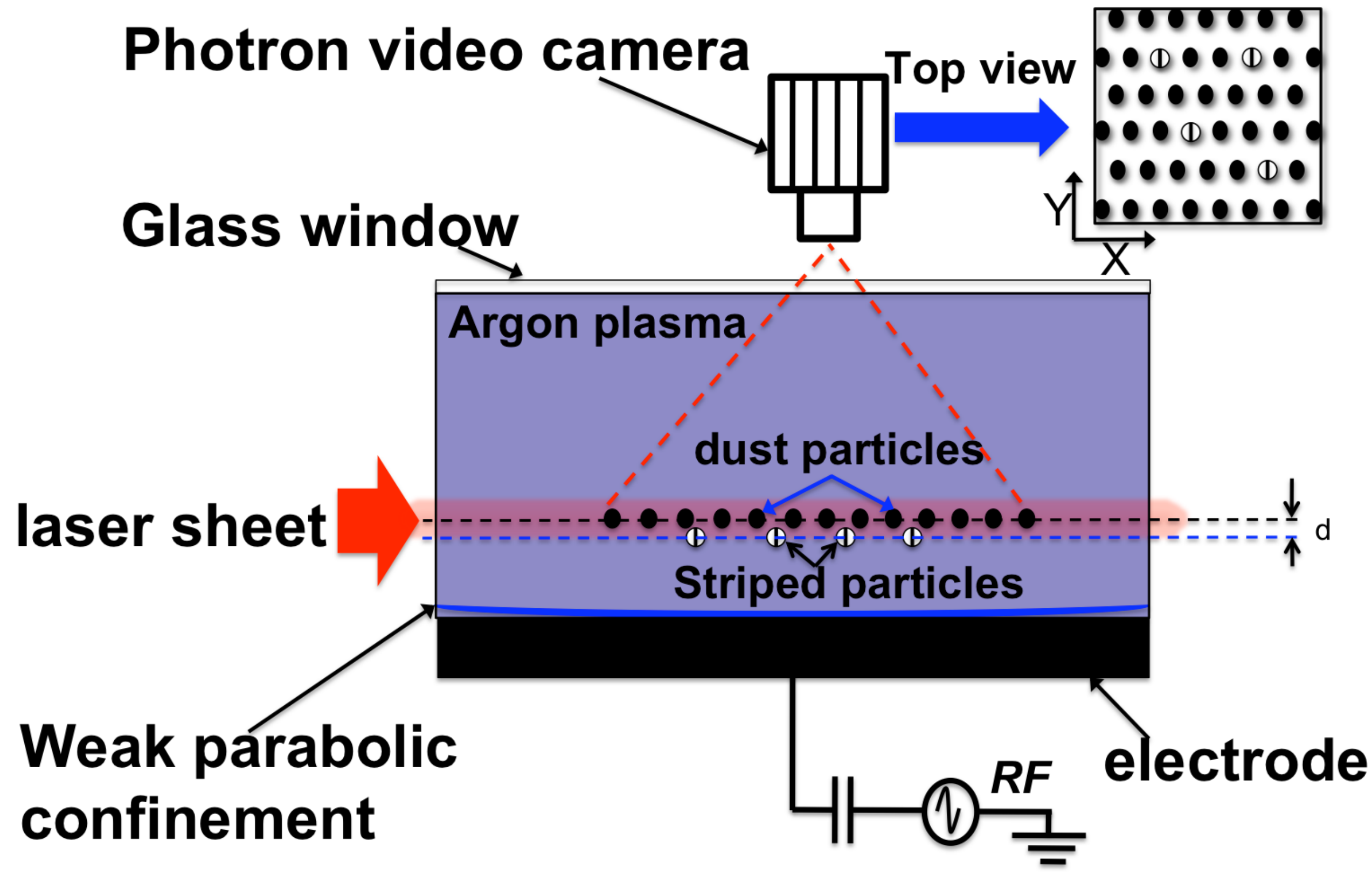}
\caption{Sketch of the experimental setup. The side view of the quasi-two-dimensional plasma crystal. Spherical microparticles (black particles) are confined in the weak parabolic confinement potential above the rf electrode and are illuminated with a horizontal laser sheet.  The binary agglomerates (striped particles) levitate just below the spherical particles without forming vertical pairs. The defocused top view image of the binary agglomerates can be identified with distinct interference fringe pattern (black stripe). Under typical experimental conditions, the vertical separation of the binary agglomerates from the monomers, d is substantially smaller than the horizontal inter-particle separation, $\Delta$.}
\label{GEC} 
\end{figure}

In this letter, we report the use of an unconventional approach with defocused imaging techniques to study quasi-two-dimensional plasma crystal (cluster) which exists before purification process. The focused image of such crystal contains significant amount of bigger (brighter) particles distributed randomly at the centre of the crystal. The presence of these particles can destroy the order completely and the system appears as disordered solid. The levitation height of the brighter particles is slightly smaller than that of the monodisperse spherical particles, but vertical pairs do not form. Defocused images of the brighter particles contain distinct interference fringe patterns which are stationary or spinning (clockwise/anticlockwise) as sketched in Fig.~\ref{GEC} and shown in Fig.~\ref{dimer_rotation}. They can be isolated visually from defocused spheres. Different confirmation tests indicate that the brighter particles with spinning interference fringe patterns on their defocused images are binary agglomerates (dimers). 
Thus defocusing technique can be used as a powerful diagnostics to identify agglomerates in complex plasmas as well as to other similar systems (ex. colloids) where particle dynamics is monitored by laser scattering methods.

\begin{figure}
\includegraphics[width=0.95\linewidth]{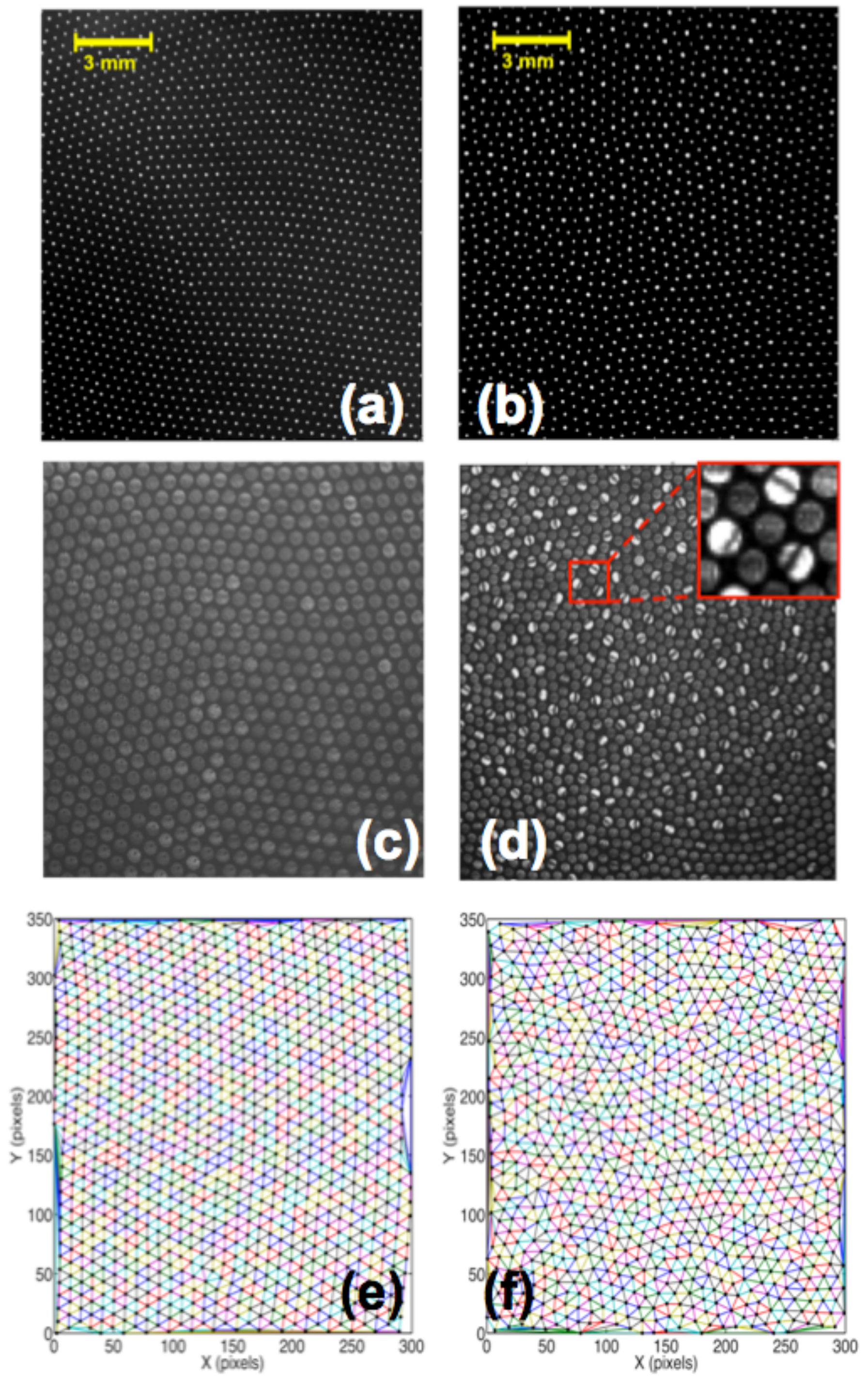}
\caption{Illustration of (a) focused and (c) defocused top-view images of two dimensional plasma crystal with monodisperse spherical particles. The quasi-two-dimensional plasma crystal with (b) focused (with brighter particles) and (d) defocused (with interference fringe patterns) top-view images are shown. 
The ``Delaunay triangulation'' for (e) two dimensional and (f) quasi-two-dimensional plasma crystal have been performed. It is clear that the presence of many brighter particles destroy the crystalline order and the system may emerge as disordered solid. In the quasi-two-dimensional plasma crystal they levitate just below the monodisperse spherical particles, but do not form vertical pairs.}
\label{image_patterns} 
\end{figure} 

The experiments were performed with a (modified) Gaseous Electronics Conference (GEC) chamber, in a capacitively coupled rf glow discharge at 13.56 MHz (see Fig.~\ref{GEC}). The Argon pressure was at 1.34 Pa and the rf power was kept fixed at 20 Watt. Melamine formaldehyde (MF) particles with a diameter ($2R$) of 7.16 $\mu$m and mass density 1.51gm/cm$^3$ were used for the experiment. The particles form a quasi-monolayer above the rf electrode with the striped particles levitating in a layer just below the monolayer. Under this experimental condition, the vertical separation of the striped particles from the monolayer is, d $\sim$ 200 $\mu$m, which is much less than the interparticle separation between the spherical particles in the monolayer ($\Delta \sim$ 600 $\mu$m). The particle suspension was illuminated with a horizontal laser sheet (wavelength of 660 nm) and imaged through the top glass window with a Photron FASTCAM 1024 PCI camera operating at a speed of 60 frames/sec with a field of view of 4.26 $\times$ 4.26 cm$^2$.
The camera lens was equipped with a narrow-band interference filter to collect only the illumination laser light scattered by the particles. Since the width of the illumination laser sheet ($\sim$ 180 $\mu$m) is less than the vertical separation of the striped particles from the monolayer, it is possible to illuminate the monolayer and the layer of striped particles at different heights separately with maximum laser intensity.

\begin{figure}
\includegraphics[width=\linewidth]{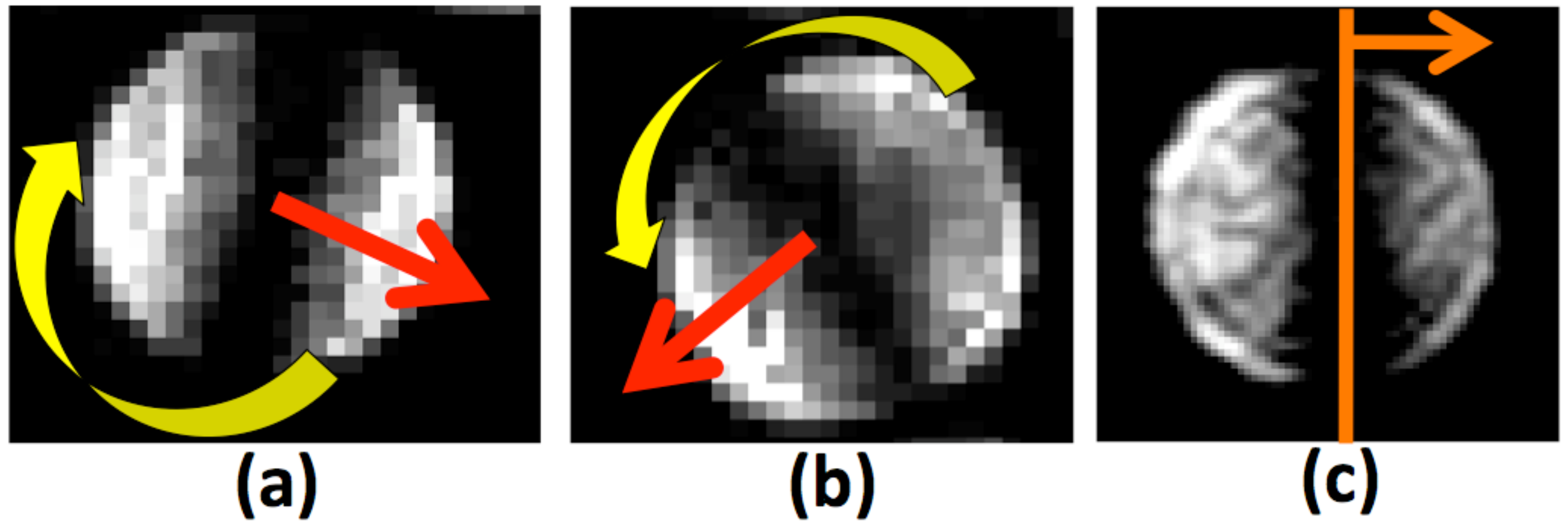}
\caption{(a) and (b) Top-view defocused images of the brighter particles which contain distinct interference fringe patterns. The interference fringe pattern can be stationary or can rotate in both clockwise/anti-clockwise directions (yellow arrow) with additional spinning motion (red arrow).\\ (c) Side-view defocused images of brighter particles with vertical interference fringe patterns. They do not rotate but move in left or right directions.}
\label{dimer_rotation}
\end{figure}

{\it Observation:}
The two dimensional plasma crystals are formed with monodisperse spherical particles after ``purification process'' through which all the irregular and bigger particles are removed. In general, the defocused images of spherical particles do not contain any interference fringes (In some situations, however, few of them contain faint stationary fringes in the direction parallel to illumination laser which may arise due to some irregularity in shape).  
The situation changes drastically before the purification process when additional brighter (bigger) particles along with monodisperse spherical particles form quasi-two-dimensional plasma crystal. The presence of brighter particles destroy the triangular crystalline order in the two dimensional plasma crystal and the system emerges as disordered solid as shown in Fig.~\ref{image_patterns}. However, the presence of few such particles maintain the order of the triangular lattice of the plasma crystal made by monodisperse spherical particles where they behave as caged particle. The brighter particles levitate just below the monodisperse spherical particles in the quasi-monolayer without forming vertical pairs. The defocused top-view images of such brighter particles contain stationary/spinning interference fringe patterns (striped particles) as shown in Fig.~\ref{dimer_rotation}. At the same time, horizontal movements of the vertically oriented fringes have been observed from the side-view images. The number of fringes increases with particle size. It is found that significant amount of such striped particles are present ($\approx 10-30 \%$) at the centre of the quasi-two-dimensional plasma crystal. 
\begin{figure}
\includegraphics[width=0.9\linewidth]{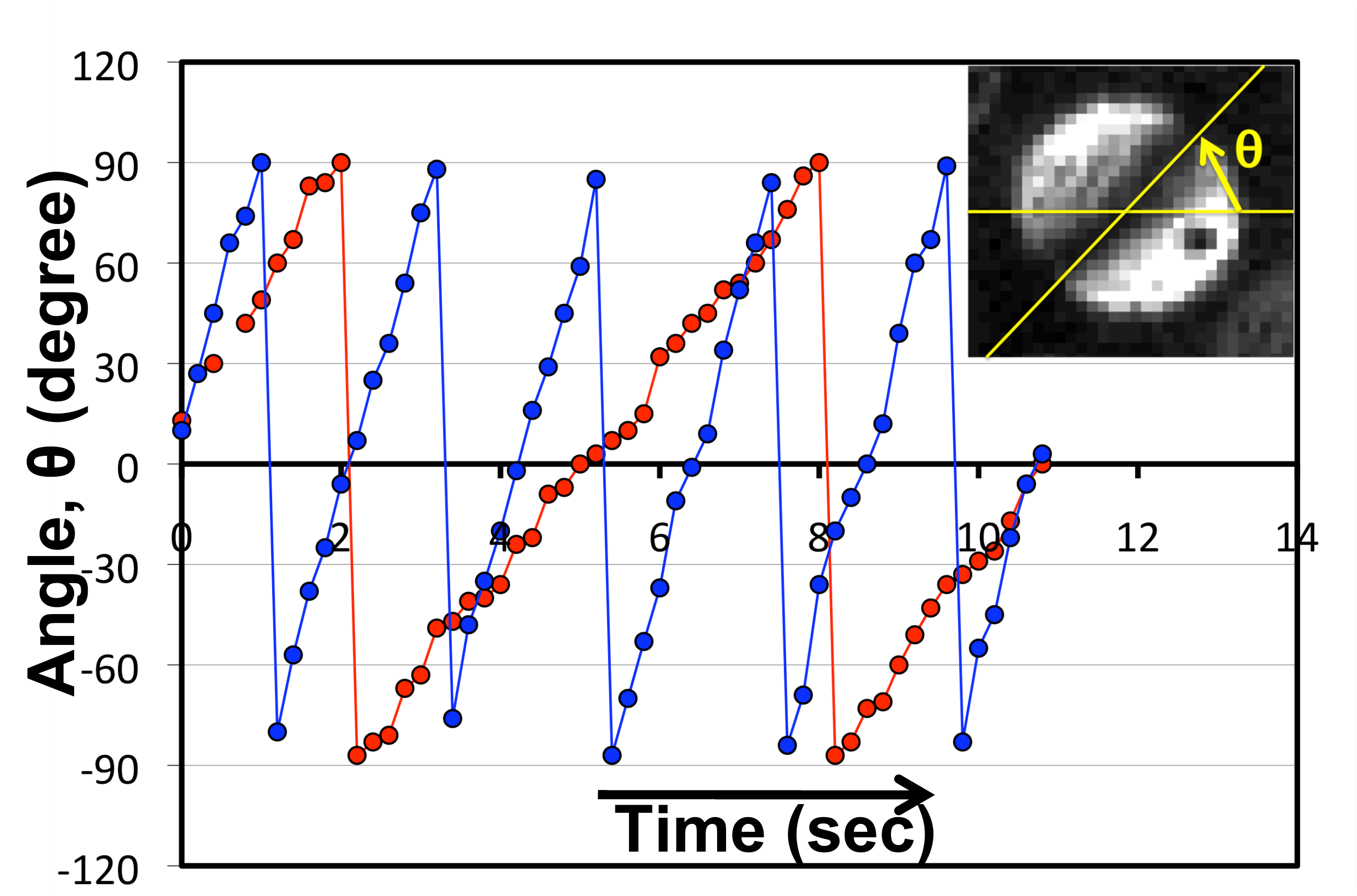}
\caption{Orientation angle of the interference fringe on the defocused images versus time. The rotational velocities are different for different particles as shown in the figure: $\sim$ 0.08 Hz (red circles) and $\sim$ 0.21 Hz (blue circles).}
\label{stripe_particles} 
\end{figure}

Three different types of striped particles have been observed: (a) the fringe movement is in the clock-wise direction, (b) the fringe movement is in the counter clock-wise direction and (c) stationary fringes with random orientation. It is found that the change of fringe orientation (angle) with time is periodic in nature. However, the rotational frequencies are different for different interference fringes as shown in Fig.~\ref{stripe_particles}. This can be due to the difference in size/shape of these particles or their positions and orientations in the sheath electric field. 
In the present configuration (Fig. 1), the angular velocity is maximum in the Y-direction, whereas it is minimum in the X-direction. The minimum rotational speed of the fringes also correspond to the change in intensity of the defocused image of the agglomerates associated with their vertical levitation height. The directional dependence of angular velocity flips when the laser illumination changes by $90^{o}$.

\begin{figure}
\includegraphics[width=\linewidth]{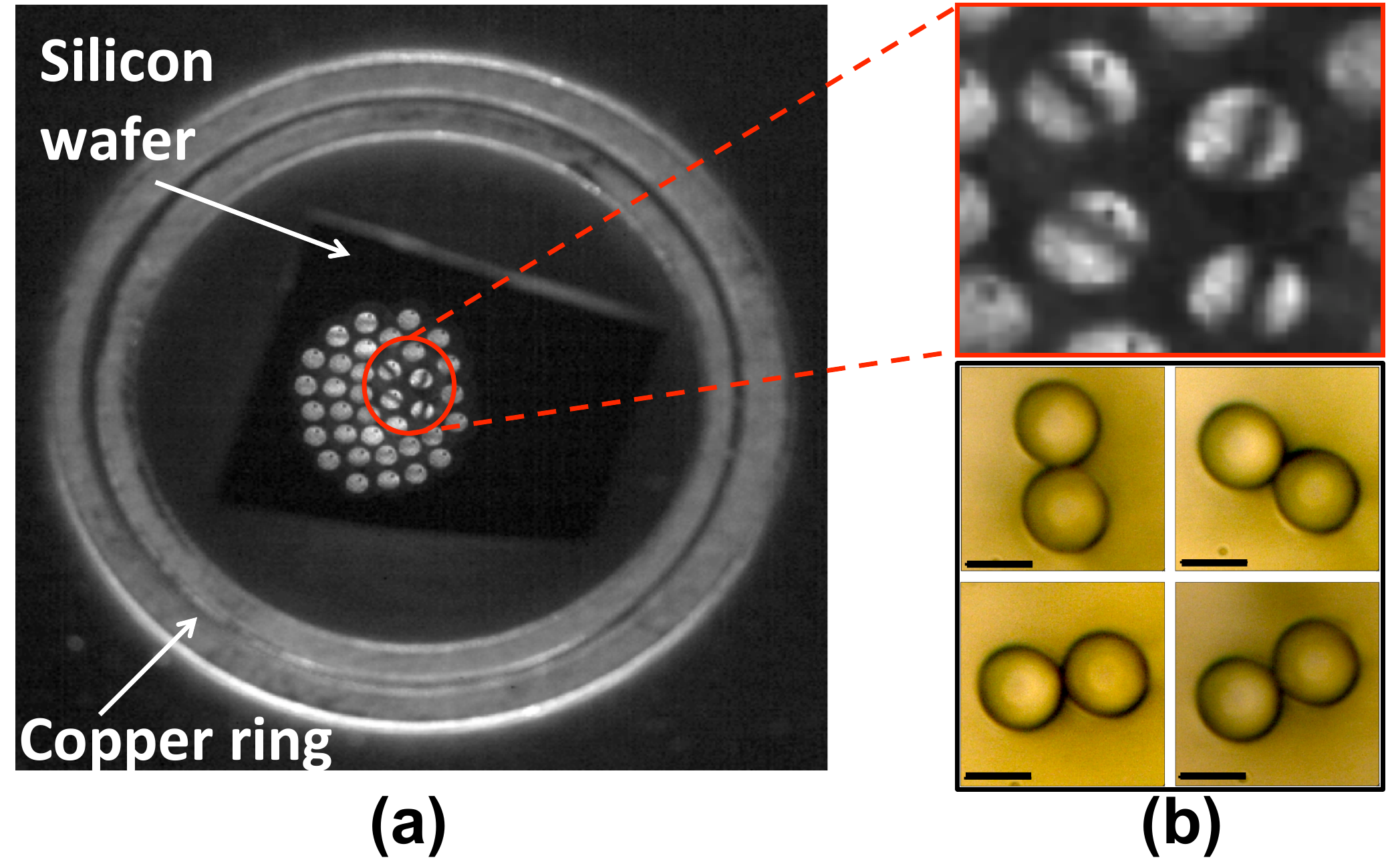}
\caption{(a) Cluster of 36 particles was formed within an additional confinement using a copper ring and a silicon wafer. The 4 striped particles were identified (inset). They dropped on the silicon wafer after the plasma was switched off. (b) The optical microscopic imaging of four binary agglomerates. The scale length (5 $\mu$m) is shown with black line. Similar results were obtained with different small clusters of particles. These observations indicate that the top-view defocused images of the particles with distinct interference fringe patterns can be binary agglomerates.
}
\label{cluster}
\end{figure}

In order to get information about these striped particles, we performed optical microscope measurements. For this purpose, we designed experiments with an additional confinement using two copper rings (each of which is of 4.8 cm in diameter and 1 mm thick). Those rings form a parabolic trap for the particles and allow the study of smaller clusters. Then, a polished silicon wafer of 400 $\mu$m thickness was placed at the centre of the copper ring which is concentric with electrode. Different small clusters were formed just above the wafer in which some particles had rotating interference fringe patterns as shown in Fig.~\ref{cluster}. Then the plasma was switched off so that all the particles could be dropped on the wafer. Afterwords, the sample (wafer with particles) was investigated using optical microscope ``Keyence VH-Z500'' with an N.A. of 0.82 (Fig.~\ref{cluster}b). All the particles of the clusters were traced. It is found that the number of binary agglomerates as traced by the optical microscope is equal to that counted as striped particles from the top-view defocused images. This suggests that the striped particles are binary agglomerates.

\begin{figure}
\includegraphics[width=\linewidth]{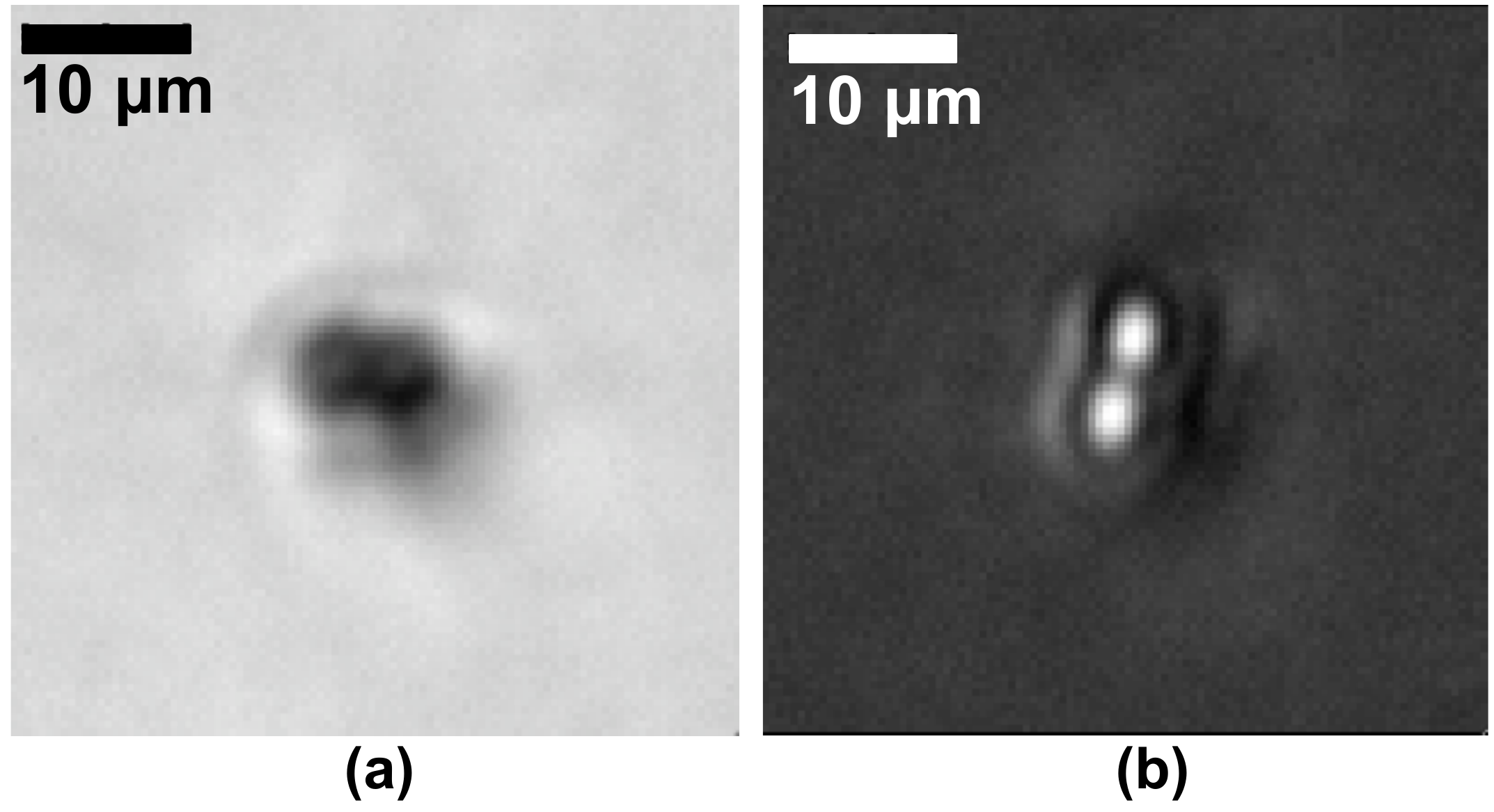}
\caption{Observation of binary agglomerates by the long-distance microscope (a) in presence of diffuse background light but in absence of laser and (b) in presence of laser. The simultaneous observations from top- and side-view were  performed.}
\label{long-dist} 
\end{figure}

To get absolute confirmation about the structure of the striped particles, we have performed similar experiments in a new plasma chamber, designed for future space laboratory for complex plasmas. In these experiments the real time dynamics of few particles was monitored by a side-view camera equiped to a ``QM 100'' long-distance microscope. The resolution of the microscope is 1.1 $\mu$m at the working distance (distance between the subject plane to the front element of the microscope) of 15 cm with N. A. of 0.142. The resolution of the microscope is good enough to resolve the shape of the used 7.16 $\mu$m particles. As mentioned before the layer of spherical particles and that of striped particles can be illuminated separately because of their vertical separation due to slightly different charge to mass ratio. It gives us the advantage to observe only the striped particles simultaneously using a top-view camera and the side-view long-distance microscope. First we monitored the particle in presence of a diffuse background light and illumination laser. Then we switched-off the illumination laser and observed the particle only against the diffuse background light with long distance microscope. This observation confirmed that we are indeed following the same particle simultaneously using top view and side view cameras. The side view images taken by the long-distance microscope confirmed that the striped particles are indeed binary agglomerates with complicated periodic dynamics. The side view images of the binary agglomerates with different orientations are shown in Fig.~\ref{long-dist}. It is found that the periodicity of their dynamics is identical to that of rotating interference fringes on its top view defocused images. The periodicity of the agglomerate dynamics from the side view images has been monitored by looking at the periodic appearance of strong light scattering by the agglomerates at a particular orientation. One such example is shown in Fig.~\ref{long-dist}. The present diagnostic systems do not allow us to resolve the correlation of the real time dynamics of binary agglomerates with the fringe orientation on its defocused top view image. It has been kept for future analysis.

Both the optical microscope and long-distance microscope measurements confirm that the striped particles are binary agglomerates which levitate below the spherical particles without forming vertical pair. They are present initially with the supplied monodisperse MF spherical particles used for the experiments and come out directly from the dispenser at the time of shaking. The spherical and striped particles can be illuminated separately due to their vertical separation. It is also observed that all the striped particles levitate at the same vertical height in the sheath above the lower electrode which indicates that all these agglomerates are of similar size, shape and charge to mass ratio (binary agglomerates).

In conclusion, we have used defocused imaging technique as a diagnostic to distinguish easily single spherical particles from agglomerates. The fact that the defocused images of binary agglomerates contain a distinct interference fringe patterns allow this discrimination. The interference fringes on the defocused images of agglomerates can be stationary or can rotate periodically. The rotational fringe pattern can be an indication of the real spinning dynamics of agglomerates. Furthermore, we used this technique to explore quasi-two-dimensional plasma crystal (cluster) consisting of monodisperse spheres and binary agglomerates. The presence of many agglomerates destroy the crystalline order made by the spherical particles and the system enters into a disordered solid state. The method can be used to explore individual asymmetric particle dynamics (orientation, spinning, precession, etc.) in unmagnetized/magnetized plasmas~\cite{Konopka_PRE.61.1890,Ishihara_2001,Tsytovich_NJP_2003,Krashennikov_POP_2010}, non-Hamiltonian dynamics, generic properties of quasi-two-dimensional plasma crystal with binary mixtures (spheres and binary agglomerates) to investigate critical point (gas-liquid phase transition), glass (jamming) transition, or to identify non-spherical particles in large 3-D complex plasmas, etc. However, all these facts are beyond the scope of our present paper and are kept for future work.

We acknowledge Dr. Satoshi Shimizu, Dr. Robert S\"utterlin and Prof. Noriyoshi Sato for helpful discussions. 

This work was supported by the German Aerospace Center (DLR)/BMWi grant no. 50WP0700

\end{document}